\documentclass[pdflatex,sn-mathphys-ay]{sn-jnl}

\usepackage{graphicx}
\usepackage{multirow}
\usepackage{amsmath,amssymb,amsfonts}
\usepackage{xcolor}
\usepackage{booktabs}
\usepackage{tabularx}
\usepackage{array}
\usepackage{rotating}
\usepackage{adjustbox}
\usepackage{bm}

\begin{document}
\title[Road-Level Heterogeneity in Crash Severity]
{Capturing Road-Level Heterogeneity in Crash Severity on Rural Two-Lane Highways: A Multilevel Statistical and Machine-Learning Analysis}

\author*[1]{\fnm{Mahdi} \sur{Azhdari}}
\email{mehdi.azhdari.1374@gmail.com}

\author[1]{\fnm{Ali Tavakoli} \sur{Kashani}}
\author[1]{\fnm{Saeideh} \sur{Amirifar}}
\author[2]{\fnm{Amirhossein} \sur{Taheri}}
\author[2]{\fnm{Gerd} \sur{M\"{u}ller}}

\affil*[1]{\orgdiv{School of Civil Engineering}, \orgname{Iran University of Science and Technology}, \orgaddress{\city{Tehran}, \country{Iran}}}

\affil[2]{\orgdiv{Department of Automotive Engineering}, \orgname{Technical University of Berlin}, \orgaddress{\city{Berlin}, \country{Germany}}}

\abstract{
Crash severity may vary across roads because crashes on the same road share contextual characteristics not fully represented by observed variables. This study examined road-level heterogeneity using 19,956 police-reported crashes from 100 rural two-lane highways in Iran during 2021--2024. Single-level logistic regression was compared with random-intercept and random-coefficient multilevel logistic models. A regression-based mixed-effects random forest (MERF) was also used for an exploratory assessment of nonlinear predictive relationships and road-level patterns.

The null multilevel model produced an intraclass correlation coefficient of 20.3\,\%, indicating meaningful latent between-road variation in crash severity. Pavement condition had the largest estimated random-slope variance, followed by lighting condition, driver education, and driver age. Logistic road-level random intercepts ranged from $-2.87$ to $+1.51$ on the log-odds scale. In a comparison excluding road-specific information, the ten-variable MERF random-forest component achieved an AUC of 0.701, compared with 0.638 for single-level logistic regression. Retaining MERF road-level corrections for roads represented during training increased AUC to 0.763. MERF road-level corrections were strongly associated with logistic random intercepts (Pearson $r=0.892$; Spearman $\rho=0.931$), indicating similar relative road patterns, although the estimates are not numerically equivalent.

The findings support complementary use of multilevel and nonlinear models. Road-level estimates may help prioritize further investigation, but they do not measure crash frequency, total road risk, or causal effects.
}

\maketitle

\section{Introduction}
\label{sec:introduction}

Road traffic crashes remain among the leading causes of death and serious injury worldwide, with a disproportionate share of fatalities occurring on rural and intercity roads in low- and middle-income countries \citep{WHO2023}. In Iran, the intercity rural road network consists largely of two-lane, two-way facilities that carry mixed traffic under limited access control. These roads combine higher operating speeds, less forgiving roadside environments, and longer emergency response times, and they account for a large share of the country's annual crash fatalities \citep{besharati2020bivariate}.

Crash severity on these roads is shaped by a wide range of factors. Driver demographics, lighting conditions, pavement quality, roadway geometry, weather, and traffic exposure all contribute to the probability that a crash results in injury rather than property damage only. Most studies of crash severity model these factors using single-level approaches in which crashes are treated as independent observations after conditioning on the included variables \citep{lord2021highway}. This framing has produced important findings about the average effects of crash-level predictors across road networks, but it overlooks a fundamental feature of crash data: individual crashes are nested within roads, and roads differ from one another in ways that cannot be fully captured by any finite set of measured variables.

These unobserved road-level differences create a hierarchical structure in crash data. Crashes on the same road share exposure levels, maintenance conditions, enforcement environment, driver population, local geometric context, and other factors that may not be directly measured. As a result, crash outcomes on the same road tend to be more similar to one another than outcomes on different roads, even after controlling for observed crash-level predictors. Standard single-level models treat this within-road correlation as residual noise. Multilevel models, by contrast, explicitly partition crash severity variation into crash-level and road-level components, making it possible to quantify how much of the total variation is attributable to road-level differences and which crash-level effects are most sensitive to local road context \citep{dupont2013multilevel,huang2010multilevel}.

This perspective has practical implications for road safety management. If a large share of crash severity variation is attributable to road-level characteristics, policies and interventions designed around average network-wide effects may be poorly calibrated. A pavement treatment or nighttime enforcement program that is effective on average across a network may have very different consequences on specific roads where the relevant risk mechanisms operate differently. Identifying which effects vary most across roads, and which roads are systematically associated with elevated risk beyond what their measured characteristics predict, can help agencies direct safety resources more precisely.

The present study addresses these questions using a hierarchical analytical framework applied to crash severity on rural two-lane roads in Iran. The primary goal is to characterize the structure of road-level heterogeneity in crash severity and to interpret what that heterogeneity means for safety management. The analysis uses a progression of multilevel logistic models to quantify between-road variation, identify roads with elevated or suppressed risk profiles, and determine which crash-level predictor effects are most context-dependent. A mixed-effects random forest is then estimated on the same predictors to compare road-level patterns across modeling frameworks, and to assess whether nonlinear mixed-effects modeling improves predictive discrimination.

The study makes two main contributions:

\begin{enumerate}
    \item \textbf{Construction of an integrated multilevel crash dataset.}
    Individual crash records were linked to road-level traffic and roadway characteristics, including AADT, access density, and heavy-vehicle ratio. This integration produced a two-level dataset of 19,956 crashes nested within 100 rural two-lane roads, allowing crash- and road-level sources of variation to be examined jointly.

    \item \textbf{Integration of multilevel statistical and mixed-effects machine-learning models.}
    Multilevel logistic models were used to quantify between-road variation, estimate road-specific effects, and examine whether selected crash-level associations varied across roads. MERF was used to assess nonlinear predictive relationships and compare the resulting road-level patterns with those obtained from the logistic framework. Together, the models provide complementary inferential and predictive perspectives on road-level heterogeneity.
\end{enumerate}

The remainder of the paper is organized as follows. Section~\ref{sec:literature} reviews related work on multilevel crash severity modeling and mixed-effects machine learning. Section~\ref{sec:data} describes the data. Section~\ref{sec:methods} presents the modeling framework. Section~\ref{sec:results} reports results with emphasis on heterogeneity findings. Section~\ref{sec:discussion} interprets the practical implications. Section~\ref{sec:conclusion} concludes.

\section{Literature Review}
\label{sec:literature}

\subsection{Multilevel Modeling and Road-Level Heterogeneity}

Researchers have become increasingly attentive to the hierarchical structure of crash data over the past two decades. Early multilevel work focused on crash frequency, showing that random-effects count models can substantially outperform their single-level counterparts by capturing unobserved segment-level or corridor-level heterogeneity \citep{elbasyouny2009accident,guo2010modeling,alarifi2018bayesian,han2018investigating,lee2018multilevel,park2020multiple}.

In crash severity modeling, hierarchical logistic and ordinal models have demonstrated similar advantages. \citet{lenguerrand2006modelling} showed that hierarchical logistic models for road crash outcomes capture between-road variance that single-level models attribute to residual noise. \citet{kim2007modeling} applied hierarchical binomial logistic models to rural intersection crash outcomes and confirmed that road-level clustering is statistically meaningful. \citet{adanu2017multilevel} identified heterogeneities in human factors across provinces, while \citet{haghighi2018impact} showed that geometric and pavement characteristics exhibit significant segment-level variance on rural two-lane highways. \citet{meng2020influential} and \citet{park2020approach} demonstrated that allowing selected effect sizes to vary across higher-level units improves crash severity predictions. More recently, \citet{alrumaidhi2024multilevel} applied a multilevel ordinal logit to over 500,000 Virginia collisions and found jurisdiction-specific shifts in crash severity patterns, with notable improvements in model fit over single-level specifications.

Multilevel models substantially improve goodness-of-fit and reveal meaningful variance components, but under fixed-effects-only prediction on held-out data, their predictive advantage over single-level models may be limited \citep{dupont2013multilevel}. This distinction between inferential and predictive value is a recurring theme in multilevel crash severity research and motivates the combined framework used in the present study.

\subsection{Machine Learning and Hybrid Modeling Approaches}

Machine-learning methods are increasingly used in crash-severity analysis because they can represent nonlinear relationships and interactions without requiring their functional forms to be specified beforehand. Their performance, however, depends on appropriate treatment of class imbalance and model evaluation. \citet{rahim2021deep} developed a deep-learning framework with a customized F1-based loss function and reported improved prediction of injury and fatal crashes, illustrating why overall accuracy alone may not adequately describe performance across severity classes. \citet{hasan2024shap} compared random forest, LightGBM, CatBoost, and XGBoost for crashes involving young drivers in New Jersey. XGBoost performed best among the evaluated models, and SHAP and sensitivity analyses were used to examine influential factors and selected combinations, including speeding under different weather conditions and multi-vehicle crashes at intersections. On mountainous two-lane roads, \citet{qi2024mountainous} compared three statistical models and six machine-learning algorithms. XGBoost achieved the strongest predictive performance in their comparison, while SHAP was used to interpret the roles of collision, roadway, vehicle, and environmental characteristics.

Other studies have combined machine-learning and statistical models to provide complementary perspectives on nonlinear relationships and unobserved heterogeneity. \citet{sun2023hybrid} combined random-forest SHAP analysis with a random-parameters logit model for crashes involving vulnerable road users. The random-parameters model was used to identify unobserved heterogeneity, while the RF-SHAP component examined interaction patterns involving variables represented by fixed and random parameters. \citet{scarano2023cyclist} similarly applied random-forest methods and a random-parameters logit model to cyclist-crash severity. The random-forest models supported prediction and variable ranking, whereas the random-parameters model identified variables exhibiting unobserved parameter heterogeneity. \citet{sadeghi2024mobility} combined XGBoost-SHAP with a random-parameters binary logit model incorporating heterogeneity in parameter means and variances to examine mobility-scooter crash severity and temporal instability. In these studies, machine learning and statistical heterogeneity models were applied as complementary components; the random-parameters statistical models were responsible for formally estimating unobserved heterogeneity.

Mixed-effects random forests provide a different approach for clustered data. The MERF framework proposed by \citet{hajjem2014mixed} combines a nonlinear random-forest component with cluster-specific random effects and was shown in simulation experiments to outperform standard random forest when random effects were non-negligible. This formulation makes it possible to represent nonlinear predictor relationships while accounting for dependence among observations belonging to the same cluster. Although mixed-effects machine-learning methods have been applied in other fields, their use in crash-severity research appears to remain limited. The present study applies MERF to crashes nested within roads and compares the relative ordering of its road-level corrections with the ordering of random intercepts obtained from a multilevel logistic model. Because these quantities are estimated on different scales and under different distributional assumptions, their magnitudes are not treated as directly comparable. This comparison evaluates whether the identified road-level injury-severity patterns remain consistent across statistical and machine-learning frameworks.

\subsection{Research gap}

Despite progress in crash-severity modeling, less work has examined multilevel heterogeneity using both statistical and machine-learning methods within a common analytical framework. Statistical multilevel models can quantify road-level variation and identify how selected predictor effects vary across roads, while mixed-effects machine learning can represent nonlinear relationships while accounting for the same hierarchical structure. Using both approaches allows their complementary capabilities to be considered and provides an opportunity to compare the road-level patterns they identify. This study applies these methods to rural two-lane highways in Iran by estimating road-level random intercepts and slopes and examining the correspondence between the relative road-level patterns obtained from multilevel logistic and mixed-effects random-forest models.
\section{Data}
\label{sec:data}

\subsection{Study area and data sources}

The analysis uses crash records from rural two-lane, two-way roads in Iran for the period 2021--2024. Crash data were obtained from police-reported incident records maintained by the Traffic Police of Iran. Each record was linked to a specific road segment using road identifiers, and only crashes on roads classified as two-lane, two-way intercity rural highways were retained. Roads with fewer than 50 crash records over the four-year period were excluded to ensure sufficient within-group observations for reliable random-effect estimation, a common criterion in multilevel crash modeling \citep{dupont2013multilevel}. The final analytical sample comprises 19,956 crash records from 100 roads.

Road-level variables were constructed from two sources. Annual average daily traffic (AADT) data were obtained from traffic monitoring stations operated by the Road Maintenance and Transportation Organization of Iran. Access density, defined as the number of entry and exit points per kilometer, was measured using high-resolution satellite imagery from Google Earth through manual identification and verification of access points along each road segment. Heavy-vehicle ratio was computed from vehicle classification counts at the same traffic monitoring stations.

\begin{figure}[htbp]
\centering
\includegraphics[width=1\textwidth]{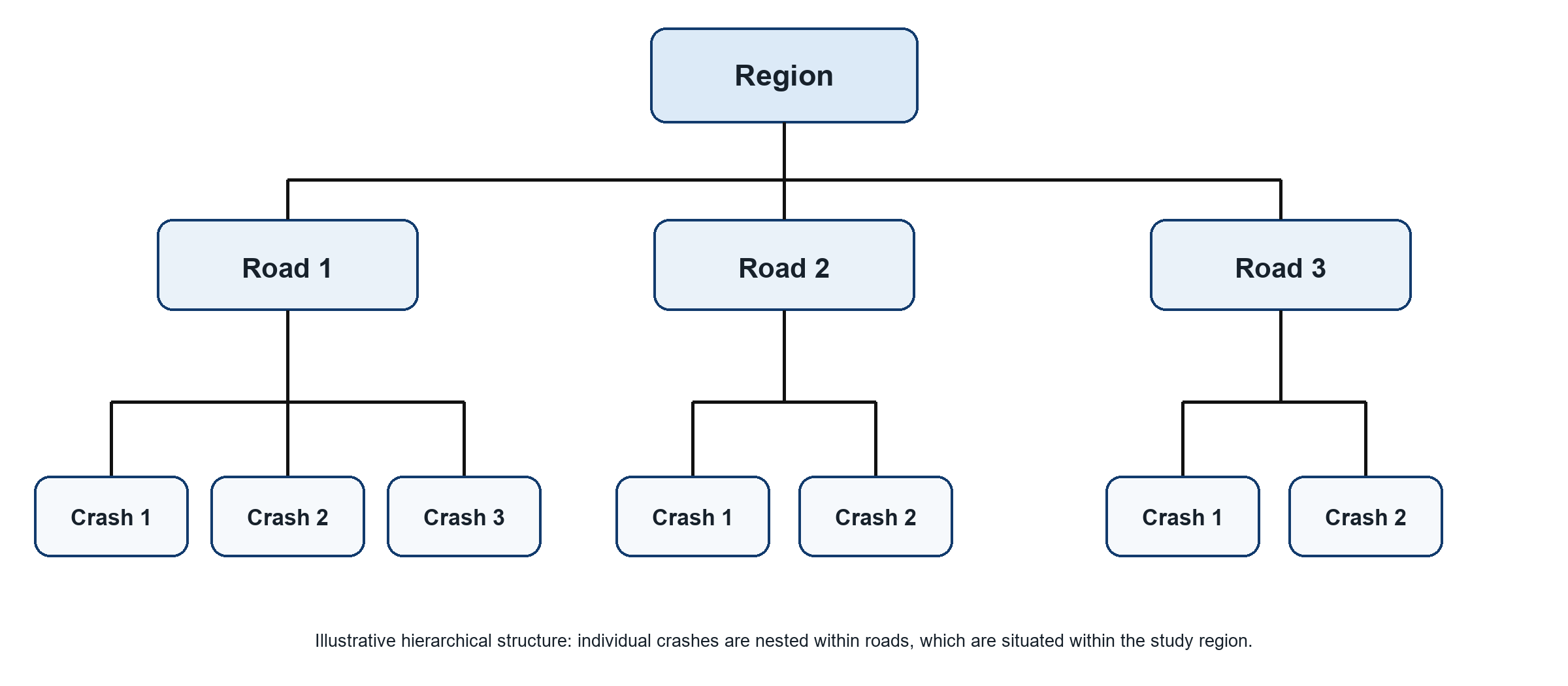}
\caption{Hierarchical structure of the crash data, with individual crash records nested within roads and predictors defined at the crash and road levels.}
\label{fig:data_hierarchy}
\end{figure}

\begin{figure}[htbp]
\centering
\includegraphics[width=1\textwidth]{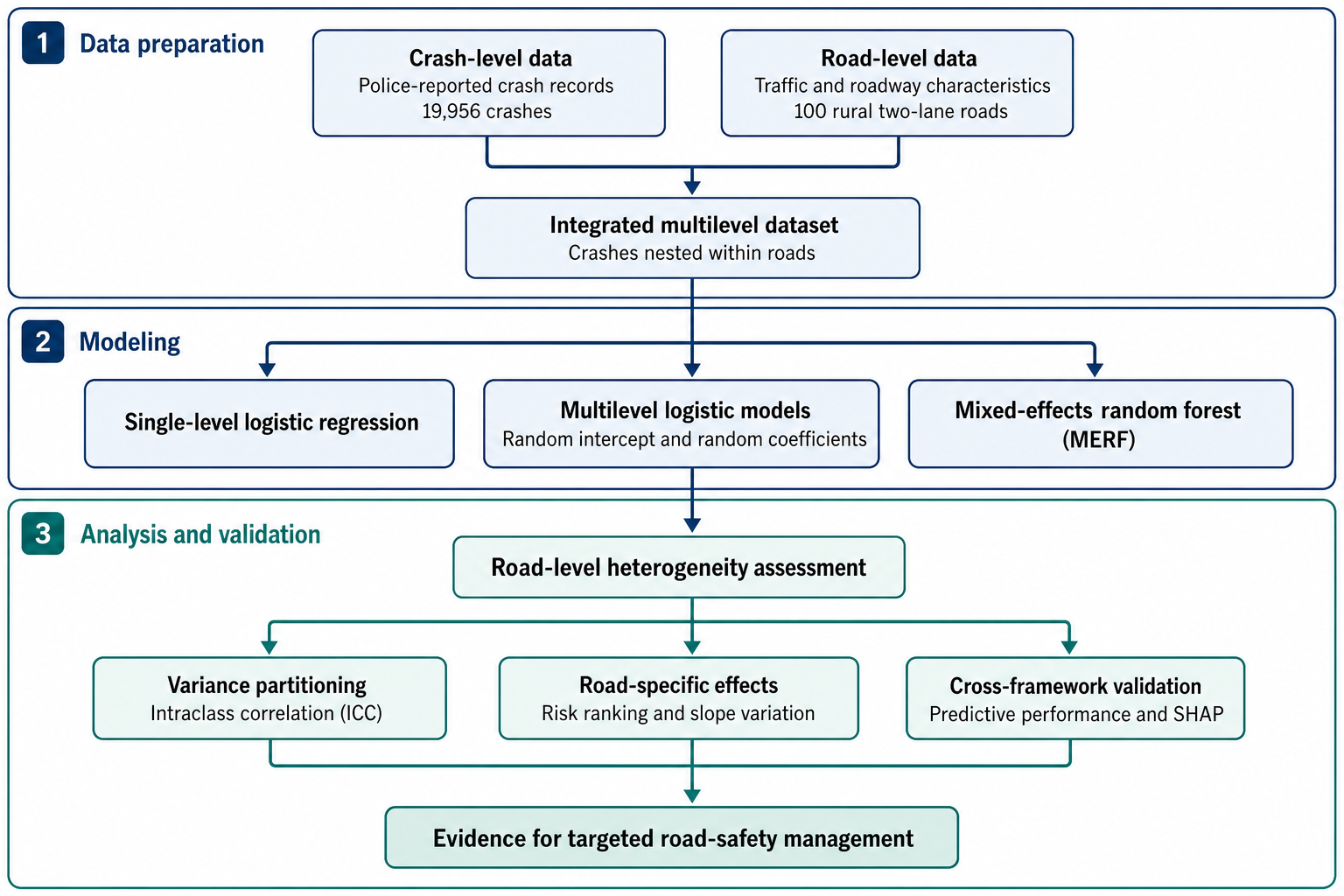}
\caption{Study workflow from crash- and road-level data collection and aggregation through model estimation and road-level heterogeneity analysis.}
\label{fig:study_workflow}
\end{figure}

The hierarchical structure of the dataset is illustrated in Figure~\ref{fig:data_hierarchy}, with individual crash records nested within roads and explanatory variables measured at both the crash and road levels. Figure~\ref{fig:study_workflow} summarizes the analytical workflow, from data collection and aggregation through estimation of the logistic and mixed-effects machine-learning models.

\subsection{Variables}

Table~\ref{tab:variables} summarizes the variables used in the analysis. The dependent variable is binary crash severity, coded as property-damage-only (PDO) versus non-PDO. Non-PDO crashes include injury and fatal crashes. This binary coding is consistent with commonly used safety analysis frameworks that distinguish crashes with human injury consequences from those involving only property damage \citep{lord2021highway}.

Crash-level predictors describe conditions at the individual crash scene. Driver education is coded as an indicator for drivers without a university degree (90.7\,\% of the sample). Driver age is coded as an indicator for drivers younger than 30 years (27.3\,\%). Driver gender is a male/female indicator (95.9\,\% male). Lighting condition indicates nighttime crashes (34.2\,\%). Pavement condition indicates adverse pavement surface conditions (11.8\,\%). Road geometry indicates crashes occurring on a vertical or horizontal curve. Weather condition indicates adverse weather.

Road-level predictors describe the broader characteristics of each road. AADT is log-transformed and mean-centered to reduce scale effects and improve numerical stability in multilevel estimation. Access density and heavy-vehicle ratio are similarly mean-centered.

\begin{table}[!htbp]
\centering
\caption{Summary of study variables}
\label{tab:variables}
\small

\begingroup
\renewcommand{\arraystretch}{1.08}
\begin{tabularx}{\linewidth}{
    @{}
    >{\raggedright\arraybackslash}l
    >{\raggedright\arraybackslash}X
    >{\centering\arraybackslash}l
    >{\raggedleft\arraybackslash}l
    @{}
}
\toprule
Variable & Description & Level & Summary \\
\midrule
Crash severity
    & PDO versus non-PDO (binary)
    & Crash
    & 40.5\,\% non-PDO \\

Driver education
    & No university degree ($1=\mathrm{yes}$)
    & Crash
    & 90.7\,\% \\

Driver age
    & Younger than 30 years ($1=\mathrm{yes}$)
    & Crash
    & 27.3\,\% \\

Driver gender
    & Male ($1=\mathrm{yes}$)
    & Crash
    & 95.9\,\% \\

Lighting condition
    & Nighttime ($1=\mathrm{yes}$)
    & Crash
    & 34.2\,\% \\

Pavement condition
    & Adverse surface ($1=\mathrm{yes}$)
    & Crash
    & 11.8\,\% \\

Road geometry
    & Curve condition ($1=\mathrm{yes}$)
    & Crash
    & 37.0\,\% \\

Weather condition
    & Adverse weather ($1=\mathrm{yes}$)
    & Crash
    & 10.0\,\% \\

AADT
    & Log-transformed and mean-centered
    & Road
    & 8,110 veh/day \\

Access density
    & Access points per kilometer, mean-centered
    & Road
    & Mean 1.09 \\

Heavy-vehicle ratio
    & Heavy-vehicle proportion, mean-centered
    & Road
    & Mean 0.104 \\
\bottomrule
\end{tabularx}
\endgroup
\end{table}
\section{Methods}
\label{sec:methods}

\subsection{Modeling framework and variable specification}

The analysis estimates six models in a structured progression. A full ten-variable single-level generalized linear model (GLM) with a logistic link provides a conventional single-level baseline using all available predictors. A reduced five-variable GLM uses only the five predictors retained in the multilevel specifications, allowing the effect of variable selection to be separated from the effect of model structure. A random-intercept multilevel logistic model and a random-coefficient multilevel logistic model both use the reduced five-variable specification, following standard practice of excluding non-significant predictors to stabilize random-effect estimation in multilevel logistic models. Finally, a mixed-effects random forest is estimated under both the reduced five-variable and full ten-variable specifications. The fitted ten-variable MERF was subsequently evaluated under two prediction conditions: complete prediction, retaining the estimated road-specific random intercepts, and fixed-effects-only prediction, suppressing those intercepts and using only the trained random-forest component. These are two evaluations of the same fitted model rather than separate model specifications.

The five-variable reduced specification retains driver education, driver age, lighting condition, pavement condition, and log-AADT. These predictors were selected because they remained stable and statistically significant across the multilevel modeling process and represent a compact combination of driver behavior, environmental condition, and road exposure factors. The five- and ten-variable MERF specifications produced similar AUC values under the selected evaluation. This suggests that adding the five supplementary predictors did not meaningfully improve predictive discrimination in this dataset.

\subsection{Single-level logistic regression}

The single-level logistic regression is:
\begin{equation}
  \mathrm{logit}(P_i) = \beta_0 + \sum_{p=1}^{P} \beta_p X_{pi},
  \label{eq:glm}
\end{equation}
where $P_i$ is the probability that crash $i$ results in non-PDO outcome, $\beta_0$ is the intercept, and $X_{pi}$ are the predictors. This model assumes all crash observations are independent conditional on the included variables, ignoring any within-road clustering.

\subsection{Null model and intraclass correlation}

Before introducing predictors, a null two-level model is estimated to measure the proportion of crash severity variation attributable to between-road differences:
\begin{equation}
  \mathrm{logit}(P_{ij}) = \beta_0 + u_{0j}, \quad u_{0j} \sim N(0, \sigma^2_{u0}),
  \label{eq:null}
\end{equation}
where $P_{ij}$ is the probability that crash $i$ on road $j$ results in non-PDO outcome and $u_{0j}$ is the road-specific random intercept. The intraclass correlation coefficient (ICC) is:
\begin{equation}
  \mathrm{ICC} = \frac{\sigma^2_{u0}}{\sigma^2_{u0} + \pi^2/3},
  \label{eq:icc}
\end{equation}
where $\pi^2/3$ approximates the level-1 variance for a logistic distribution. An ICC above 5\,\% is commonly used as evidence that clustering should be accounted for in the analysis \citep{dupont2013multilevel}.

\subsection{Random-intercept multilevel logistic model}

The random-intercept model adds crash-level and road-level predictors while allowing each road to have its own baseline log-odds:
\begin{equation}
  \mathrm{logit}(P_{ij}) = \beta_0 + \sum_{p=1}^{P} \beta_p X_{pij} + u_{0j}.
  \label{eq:ri}
\end{equation}
The random intercept $u_{0j}$ captures unobserved road-specific baseline risk. A road with a large positive intercept is systematically more dangerous than its observed characteristics predict; a road with a large negative intercept is systematically safer. These road-level estimates are the primary tool for identifying roads that warrant targeted site investigation.

\subsection{Random-coefficient multilevel logistic model}

The random-coefficient model extends the random-intercept specification by allowing selected crash-level predictor effects to vary across roads:
\begin{equation}
  \mathrm{logit}(P_{ij}) = \beta_0 + \sum_{p=1}^{P} \beta_p X_{pij} + u_{0j} + \sum_{k=1}^{K} u_{kj} X_{kij},
  \label{eq:rc}
\end{equation}
where $u_{kj}$ is the road-specific random slope for predictor $X_{kij}$. Random slopes are estimated for driver education, driver age, lighting condition, and pavement condition. The magnitude of $\mathrm{Var}(u_{kj})$ indicates how much the effect of predictor $k$ varies across roads: a larger variance means the predictor has more context-dependent consequences. Log-AADT is retained as a fixed effect because it is a road-level characteristic that does not vary within roads and therefore cannot carry a meaningful random slope in a crash-level analysis.

\subsection{Mixed-effects random forest}

The MERF model extends the logistic framework by replacing the linear fixed-effects component with a nonlinear random forest:
\begin{equation}
  y_{ij} = f(X_{ij}) + Z_{ij} b_j + \varepsilon_{ij},
  \label{eq:merf}
\end{equation}
where $f(X_{ij})$ is a random forest function, $Z_{ij} b_j$ is the road-level random effect, and $\varepsilon_{ij}$ is residual error. MERF is estimated using an expectation-maximization (EM) algorithm that alternates between fitting the random forest to residuals adjusted for current random-effect estimates and updating the random effects given current forest predictions \citep{hajjem2014mixed}. The available MERF implementation is formulated as a regression model with Gaussian residual errors. In this study, the binary outcome was coded as 0 and 1 and analyzed using this regression-based formulation. The resulting outputs are therefore treated as predictive scores rather than calibrated probabilities. A monotonic sigmoid transformation was applied for threshold-based classification, while AUC was unaffected by this transformation. Accordingly, the MERF analysis is interpreted as an exploratory nonlinear mixed-effects comparison rather than as a binary mixed-effects model equivalent to multilevel logistic regression.

Both models produce road-level adjustments, but the estimates are not directly comparable in magnitude because they are defined on different response scales and under different distributional assumptions. The comparison, therefore, focuses on their association and the relative ordering of roads. Their correlation with the logistic random intercepts is used to assess whether the two frameworks identify similar road-level risk patterns, providing a cross-framework comparison of the heterogeneity findings. SHAP (SHapley Additive exPlanations) values computed from the random forest component provide insight into the relative importance of each predictor and their distributional effects, complementing the coefficient-based interpretation from the logistic models.

\subsection{Model estimation and evaluation}

All models were estimated using an 80\,\% random training split of the 19,956 crash records, with predictive performance evaluated on the remaining 20\,\%. Because the split was performed at the crash level, roads represented in the test set could also be represented in the training data. Classification thresholds were selected by maximizing the F1-score using the training data and were then applied to the test set without adjustment. AUC was treated as the primary comparison metric because it is independent of the selected classification threshold.

For the multilevel logistic models, test-set predictions were generated using only the fixed-effects component, with road-specific random effects suppressed. This provides a conservative, road-agnostic evaluation of each model's predictor coefficients and is not directly comparable to MERF's known-road (full) prediction condition, which explicitly retains road-level corrections for roads represented during training. MERF was evaluated under two conditions. In the first, road-specific random intercepts were retained for test crashes occurring on roads represented during training, providing a conditional known-road prediction. In the second, these intercepts were suppressed, and predictions were generated using only the trained random-forest component.

Model fit was assessed using deviance ($-2\log L$) and the Akaike Information Criterion (AIC). Multilevel logistic models were estimated in R using \texttt{glmer} from the \texttt{lme4} package with the Laplace approximation and the BOBYQA optimizer. MERF was estimated in Python using the \texttt{merf} package with a random forest regressor base learner and road ID as the grouping variable.

\section{Results}
\label{sec:results}

\subsection{Between-Road Variance and Intraclass Correlation}
\label{subsec:icc}

The null model provides the foundational evidence for road-level heterogeneity. The estimated between-road variance is $\hat{\sigma}^2_{u0} = 0.8375$, and the binary logistic level-1 variance is $\pi^2/3 = 3.2899$. The ICC is therefore:
\[
  \mathrm{ICC} = \frac{0.8375}{0.8375 + 3.2899} = 0.203 \quad (20.3\,\%).
\]
This value indicates that approximately one-fifth of the latent variation in crash severity occurred between roads before explanatory variables were introduced. The ICC exceeds the 5\,\% threshold commonly used to support accounting for clustering in multilevel analysis \citep{dupont2013multilevel}. This result indicates meaningful between-road heterogeneity in the null model but does not identify the characteristics responsible for that variation.

Table~\ref{tab:icc} summarizes the null-model variance components, the resulting ICC, and the analytical sample.

\begin{table}[!htbp]
\centering
\caption{Null-model variance components and intraclass correlation coefficient}
\label{tab:icc}
\small

\begingroup
\renewcommand{\arraystretch}{1.12}
\begin{tabularx}{\linewidth}{
    @{}
    >{\raggedright\arraybackslash}X
    >{\centering\arraybackslash}l
    >{\raggedright\arraybackslash}X
    @{}
}
\toprule
Metric & Value & Interpretation \\
\midrule

Between-road variance
($\hat{\sigma}^{2}_{u0}$)
    & 0.8375
    & Latent variance attributable to differences between roads \\

Level-1 variance
($\pi^{2}/3$)
    & 3.2899
    & Assumed individual-level variance for the logistic model \\

Intraclass correlation coefficient
    & 20.3\,\%
    & Share of latent crash-severity variation occurring between roads \\

Reference threshold
    & 5.0\,\%
    & Common threshold supporting the use of multilevel modeling \\

Analytical sample
    & 19,956 crashes
    & Crash records nested within 100 roads \\
\bottomrule
\end{tabularx}
\endgroup
\end{table}

\subsection{Model coefficients and goodness of fit}
\label{subsec:coefficients}

Table~\ref{tab:coefficients} presents the estimated coefficients and goodness-of-fit statistics. The full GLM uses all ten predictors; the reduced GLM and the multilevel logistic models use the five-variable specification.

\begin{sidewaystable}
\caption{Coefficients and goodness-of-fit for logistic models\label{tab:coefficients}}
\small
\begin{tabular*}{\textheight}{@{\extracolsep\fill}lrrrr}
\toprule
Variable & GLM full (10 vars) & GLM reduced (5 vars) & Random intercept & Random coefficient \\
\midrule
\multicolumn{5}{l}{\textit{Fixed effects (log-odds coefficients)}} \\
Intercept           & $-0.635^*$ & $-0.761^*$ & $-0.655^*$ & $-0.643^*$ \\
Driver education    & $0.355^*$  & $0.325^*$  & $0.455^*$  & $0.445^*$  \\
Driver age ($<$30)  & $0.208^*$  & $0.222^*$  & $0.243^*$  & $0.252^*$  \\
Lighting (night)    & $0.205^*$  & $0.186^*$  & $0.157^*$  & $0.151^*$  \\
Pavement (adverse)  & $-0.365^*$ & $-0.565^*$ & $-0.323^*$ & $-0.345^*$ \\
Driver gender       & $-0.137$   & ---        & ---        & ---        \\
Weather (adverse)   & $-0.209$   & ---        & ---        & ---        \\
Road geometry       & $-0.092^*$ & ---        & ---        & ---        \\
AADT (log, centered)& $-1.254^*$ & $-1.130^*$ & $-0.562^*$ & $-0.562^*$ \\
Access density      & $0.300^*$  & ---        & ---        & ---        \\
Heavy-vehicle ratio & $-0.605^*$ & ---        & ---        & ---        \\
\midrule
\multicolumn{5}{l}{\textit{Random effects (variance components)}} \\
Intercept variance  & ---  & ---  & 0.782 & 0.650 \\
Driver education    & ---  & ---  & ---   & 0.013 \\
Driver age          & ---  & ---  & ---   & 0.009 \\
Lighting condition  & ---  & ---  & ---   & 0.017 \\
Pavement condition  & ---  & ---  & ---   & 0.048 \\
\midrule
\multicolumn{5}{l}{\textit{Goodness of fit (training split)}} \\
Deviance ($-2LL$)   & 20,492 & 20,688 & 17,733 & 17,725 \\
AIC                 & 20,514 & 20,700 & 17,747 & 17,767 \\
\bottomrule
\end{tabular*}
\footnotetext{$^*p<0.05$. Full GLM uses all 10 predictors. Reduced GLM, random-intercept, and random-coefficient models use the same 5 predictors.}
\end{sidewaystable}

The coefficient signs are consistent across logistic specifications. Lack of university education was associated with a higher probability of a non-PDO outcome. Educational attainment may capture differences in risk perception, restraint use, driving exposure, or other behavioral and socioeconomic characteristics that were not directly measured. Younger drivers ($<$30 years) show higher severity risk. Nighttime conditions are associated with higher injury probability. Adverse pavement condition carries a negative fixed coefficient across all multilevel specifications. This sign does not imply that poor pavement is generally protective; rather, it likely reflects behavioral adaptation in the average road in this sample, where drivers may reduce speed under visibly poor surface conditions. The substantial random-slope variance for pavement (discussed below) indicates that this average effect conceals important road-specific variation.

The pattern of model fit is consistent with the hierarchical structure confirmed by the ICC. Deviance decreases from 20,688 for the reduced GLM to 17,733 for the random-intercept model, a reduction of 2,955 units that indicates substantial unexplained road-level structure in the single-level model. The random-coefficient model produces a further modest reduction to 17,725, with additional insight into slope variation discussed in the next section.

\subsection{Road-Specific Variation in Crash-Level Effects}
\label{subsec:slopes}

The random-coefficient model produced nonzero slope-variance estimates for all four tested crash-level predictors, although their magnitudes differed. Table~\ref{tab:randomslopes} presents the random-slope variance estimates.

\begin{table}[htbp]
\centering
\caption{Random-slope variance estimates from the random-coefficient model\label{tab:randomslopes}}
\small
\begin{tabular}{@{}lc@{}}
\toprule
Variable & Variance \\
\midrule
Pavement condition (adverse) & 0.048 \\
Lighting condition (night)   & 0.017 \\
Driver education             & 0.013 \\
Driver age ($<$30)           & 0.009 \\
Intercept                    & 0.650 \\
\bottomrule
\end{tabular}
\end{table}

Pavement condition shows the largest random-slope variance among the four tested predictors. This is the most practically important finding of the random-coefficient analysis. On some roads adverse pavement is associated with substantially lower injury probability, consistent with drivers reducing speed when they perceive poor surface conditions. On other roads the same pavement condition has a weaker or different relationship with severity, possibly because traffic speeds, road width, sight distance, or other contextual factors modify the behavioral response. The large slope variance for pavement means that the negative average coefficient reported in Table~\ref{tab:coefficients} should not be interpreted as a uniform network-wide effect. The safety consequences of adverse pavement are road-specific, and pavement-related safety interventions need to account for local context rather than relying on a single average estimate.

Lighting condition shows the second largest slope variance. The relationship between nighttime driving and crash severity is moderated by road-level factors such as street lighting infrastructure, speed limits, traffic density, and the typical driver population on each road. Roads where nighttime conditions are consistently associated with elevated injury risk are candidates for lighting upgrades or enforcement during dark hours. Roads where the nighttime effect is minimal or reversed may reflect different operating conditions that reduce the role of lighting in crash outcomes.

\subsection{Distribution of Road-Level Random Intercepts}
\label{subsec:roadranks}

\begin{figure}[htbp]
  \centering
  \includegraphics[width=\textwidth]{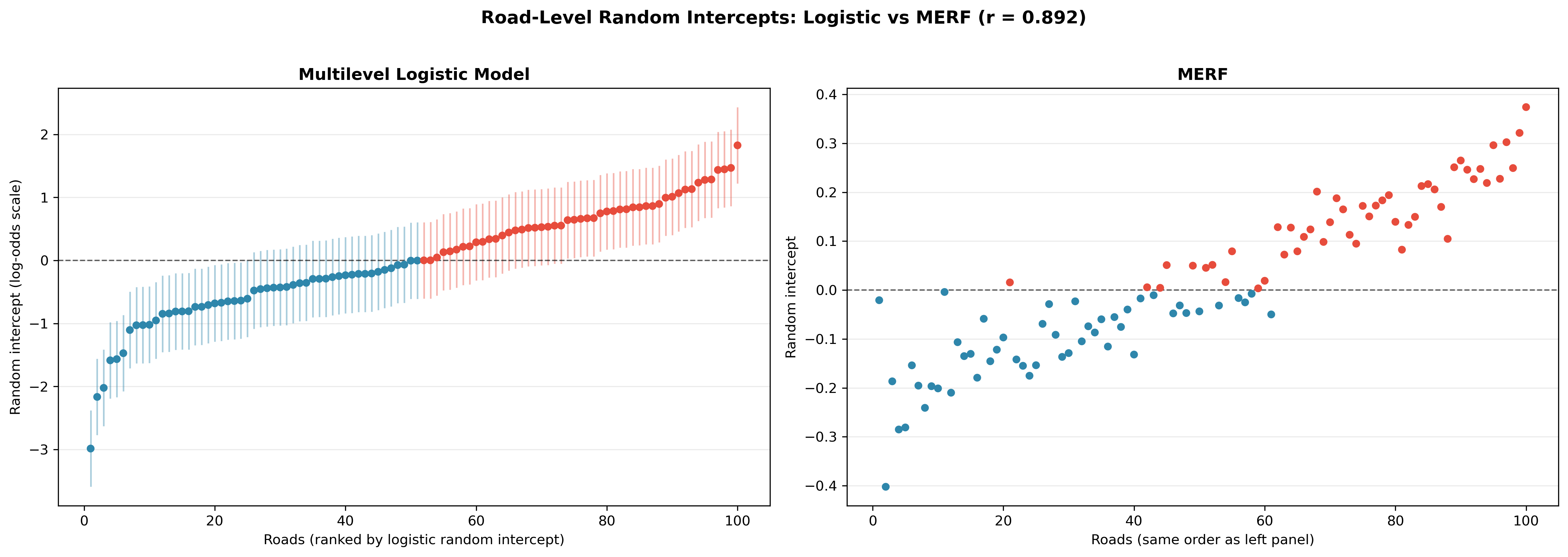}
  \caption{Road-level random intercepts from the multilevel logistic model (left) and MERF (right), ranked identically for comparison ($r = 0.892$).}
  \label{fig:caterpillar}
\end{figure}

The random-intercept model provides road-specific baseline risk estimates after controlling for all measured crash-level and road-level predictors. Figure~\ref{fig:caterpillar} shows all 100 roads ranked by their estimated random intercept with 95\,\% confidence intervals. The substantial spread of intercepts, ranging from $-2.87$ to $+1.51$ on the log-odds scale, confirms that road identity is a meaningful predictor of crash severity beyond the included covariates.

Table~\ref{tab:roadrankings} lists the ten roads with the highest and lowest estimated random intercepts. Higher positive values indicate greater conditional propensity for a recorded crash to result in a non-PDO outcome, whereas more negative values indicate lower conditional injury-severity propensity after accounting for the observed crash- and road-level predictors.

\begin{table*}[!htbp]
\centering
\caption{Roads with the highest and lowest conditional injury-severity propensity based on random-intercept estimates\label{tab:roadrankings}}
\small

\begin{adjustbox}{max width=\linewidth}
\begin{tabular}{@{}r l r l r@{}}
\toprule
Rank &
\multicolumn{1}{c}{Higher-injury-severity} &
\multicolumn{1}{c}{Random} &
\multicolumn{1}{c}{Lower-injury-severity} &
\multicolumn{1}{c}{Random} \\
&
\multicolumn{1}{c}{road ID} &
\multicolumn{1}{c}{intercept} &
\multicolumn{1}{c}{road ID} &
\multicolumn{1}{c}{intercept} \\
\midrule
1  & 79  &  1.505 & 96  & -2.866 \\
2  & 56  &  1.482 & 86  & -2.272 \\
3  & 80  &  1.469 & 44  & -2.017 \\
4  & 5   &  1.425 & 39  & -1.534 \\
5  & 122 &  1.245 & 47  & -1.484 \\
6  & 125 &  1.186 & 46  & -1.469 \\
7  & 116 &  1.130 & 37  & -1.007 \\
8  & 3   &  0.989 & 127 & -0.892 \\
9  & 23  &  0.969 & 64  & -0.877 \\
10 & 115 &  0.939 & 27  & -0.849 \\
\bottomrule
\end{tabular}
\end{adjustbox}

\parbox{0.95\linewidth}{\footnotesize\textit{Note:} Random intercepts are expressed on the log-odds scale and describe residual differences in the probability that a recorded crash results in a non-PDO outcome, conditional on the included predictors.}
\end{table*}

\begin{figure}[htbp]
  \centering
  \includegraphics[width=0.75\textwidth]{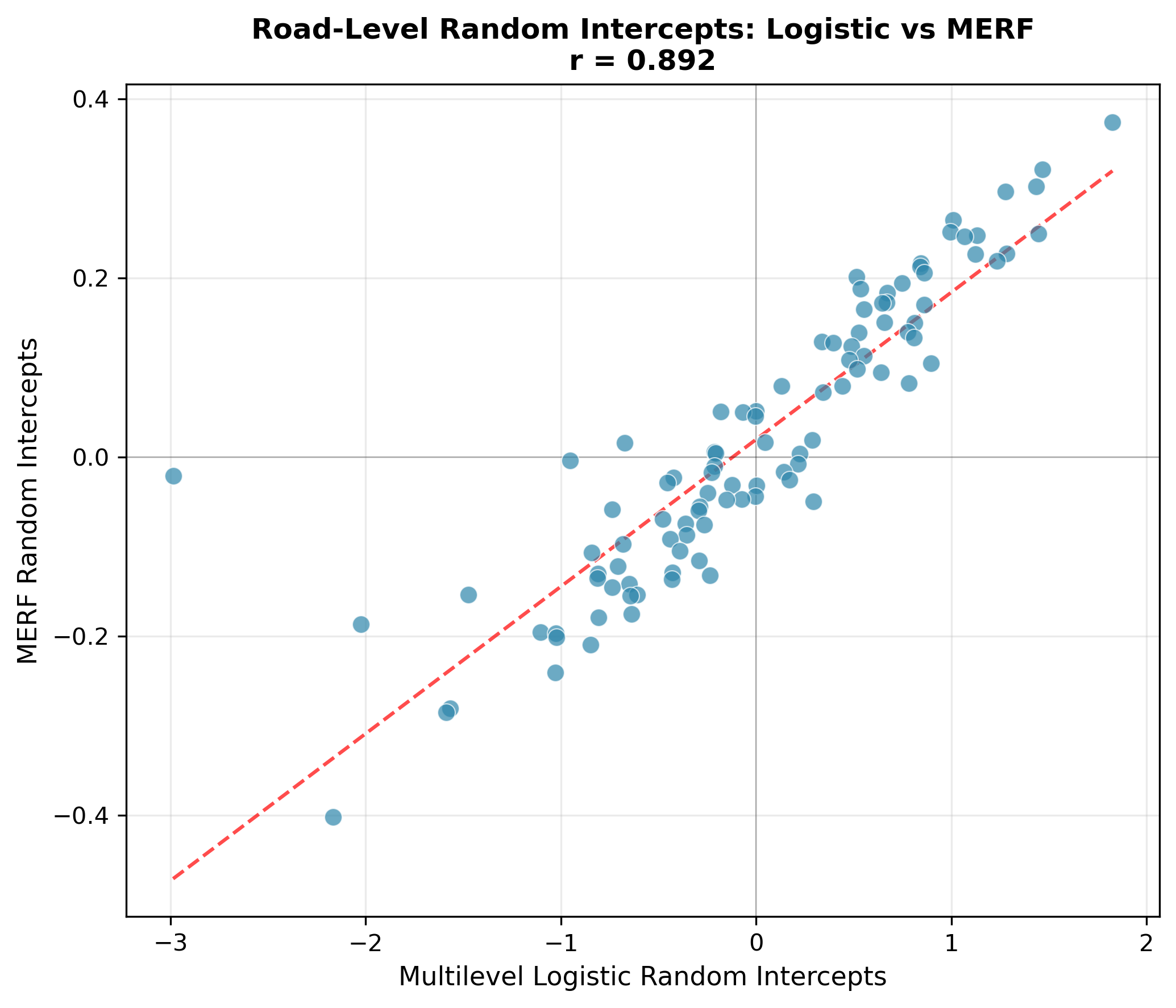}
\caption{Association between multilevel logistic random intercepts and MERF road-level corrections. Each point represents one road. Because the estimates are defined on different scales, the comparison concerns their association and relative ordering rather than numerical equivalence. Pearson $r=0.892$ (95\,\% CI: 0.844--0.926, $p<0.001$); Spearman $\rho=0.931$ ($p<0.001$).}
  \label{fig:ri_comparison}
\end{figure}

Figure~\ref{fig:ri_comparison} compares the road-level random intercepts estimated by the multilevel logistic model and MERF. The estimates were strongly associated across the two frameworks (Pearson $r=0.892$, 95\,\% CI: 0.844--0.926, $p<0.001$). The corresponding Spearman rank correlation was $\rho=0.931$ ($p<0.001$), generally indicating similar road rankings.

The strong Pearson and Spearman correlations indicate broadly similar road-level patterns across the two frameworks. This agreement suggests that the relative ordering of roads is not highly sensitive to the use of a logistic or random-forest fixed-effects component. Despite their different formulations, the two models produced similar relative rankings of the roads represented in the analysis.

\subsection{SHAP analysis of MERF}
\label{subsec:shap}

\begin{figure}[htbp]
  \centering
  \includegraphics[width=0.85\textwidth]{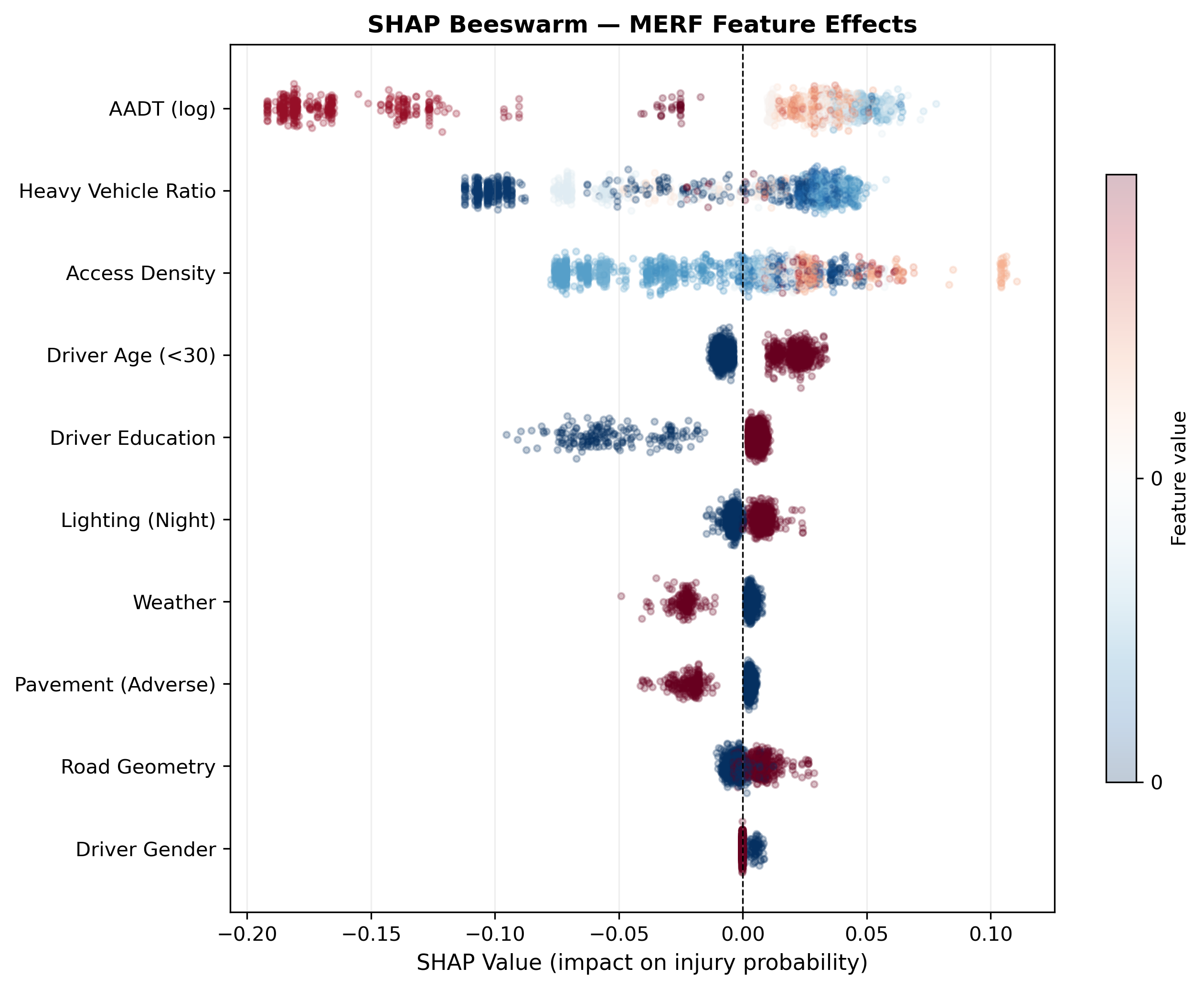}
\caption{SHAP values for the ten-variable MERF model. Position indicates each predictor's contribution, and color indicates its feature value (blue = low; red = high).}
  \label{fig:shap_beeswarm}
\end{figure}

Figure~\ref{fig:shap_beeswarm} shows SHAP values calculated from the random-forest component of the ten-variable MERF specification. The values describe each predictor's contribution to the model predictions.

AADT had the largest mean absolute SHAP value, indicating the largest average contribution to predictions from the MERF fixed-effects component. Higher traffic volumes were generally associated with lower injury probability, which may partly reflect lower operating speeds on more heavily traveled roads. Driver education, driver age, pavement condition, and lighting condition also show meaningful SHAP contributions, with directional effects consistent with the logistic coefficient signs. The five supplementary predictors generally had smaller SHAP contributions than the leading predictors. This pattern is consistent with the five- and ten-variable MERF specifications not meaningfully differing in predictive discrimination.

\subsection{Predictive performance}
\label{subsec:performance}

Table~\ref{tab:performance} presents held-out test-set performance across the six fitted models and the additional fixed-effects-only evaluation of the ten-variable MERF. Classification thresholds were selected separately for each model or prediction condition by maximizing F1 on the training data and were then applied unchanged to the test data. The full GLM uses ten predictors, whereas the reduced GLM and both multilevel logistic models use the five-variable specification. The fixed-effects-only and full ten-variable MERF results are two prediction conditions for the same fitted model, not separate models. The former suppresses road-specific random intercepts, while the latter retains them for test crashes on roads represented during training.

\begin{table*}[!htbp]
\centering
\caption{Held-out test-set performance across model specifications. Classification thresholds were selected using the training data only. AUC is the primary threshold-independent comparison metric.\label{tab:performance}}
\small

\begin{adjustbox}{max width=\linewidth}
\begin{tabular}{@{}lrrrrrrr@{}}
\toprule
Model & Vars & Threshold & Accuracy & Precision & Recall & F1 & AUC \\
\midrule
GLM full
    & 10 & 0.286 & 0.546 & 0.470 & 0.916 & 0.622 & 0.638 \\
GLM reduced
    & 5 & 0.31 & 0.552 & 0.470 & 0.887 & 0.615 & 0.618 \\
Random intercept
    & 5 & 0.39 & 0.517 & 0.450 & 0.899 & 0.600 & 0.616 \\
Random coefficient
    & 5 & 0.39 & 0.519 & 0.451 & 0.899 & 0.601 & 0.616 \\
MERF (5 variables), full prediction
    & 5 & 0.59 & 0.672 & 0.568 & 0.812 & 0.668 & 0.763 \\
MERF (10 variables), fixed-effects-only
    & 10 & 0.596 & 0.603 & 0.507 & 0.904 & 0.650 & 0.701 \\
MERF (10 variables), full prediction
    & 10 & 0.596 & 0.673 & 0.569 & 0.804 & 0.667 & 0.763 \\
\bottomrule
\end{tabular}
\end{adjustbox}
\end{table*}

Three patterns in Table~\ref{tab:performance} are worth noting. First, reducing the GLM from ten to five predictors produced a relatively small decline in AUC, from 0.638 to 0.618, suggesting that the five retained predictors captured much of the discrimination available to the logistic specification. Second, the random-intercept and random-coefficient logistic models produced AUC values of 0.616 under fixed-effects-only prediction — nearly identical to the reduced GLM's 0.618. This is expected, since suppressing the random effects removes the road-specific information that would otherwise differentiate these models from the single-level GLM at prediction time. The similar AUC values should therefore not be interpreted as evidence that accounting for road-level heterogeneity is unnecessary: the ICC (20.3\,\%) and the large deviance reduction from adding random intercepts (Table~\ref{tab:coefficients}) already demonstrate that road identity carries substantial explanatory value — it is simply not exercised under this fixed-effects-only evaluation condition.

Third, in the matched ten-variable comparison without road-specific information, the MERF fixed-effects component achieved an AUC of 0.701, compared with 0.638 for the GLM. The difference of 0.063 shows that the MERF random-forest component provided greater discrimination than the logistic specification when neither prediction included road-specific information. This difference may reflect nonlinearities and interactions captured by the forest, but it cannot be attributed exclusively to functional-form flexibility because the models also differ in their estimation and distributional assumptions. Retaining MERF's road-specific random intercepts increased AUC by a further 0.062, to 0.763, for crashes on roads represented during training. The full MERF advantage should therefore not be attributed entirely to the nonlinear functional form. Of the total AUC difference between the full MERF and GLM results, 0.063 was already present when road-specific information was excluded, while an additional 0.062 was observed when MERF road-level corrections were retained for known roads. The five- and ten-variable full MERF specifications did not meaningfully differ in AUC.

\begin{figure}[htbp]
  \centering
  \includegraphics[width=0.85\textwidth]{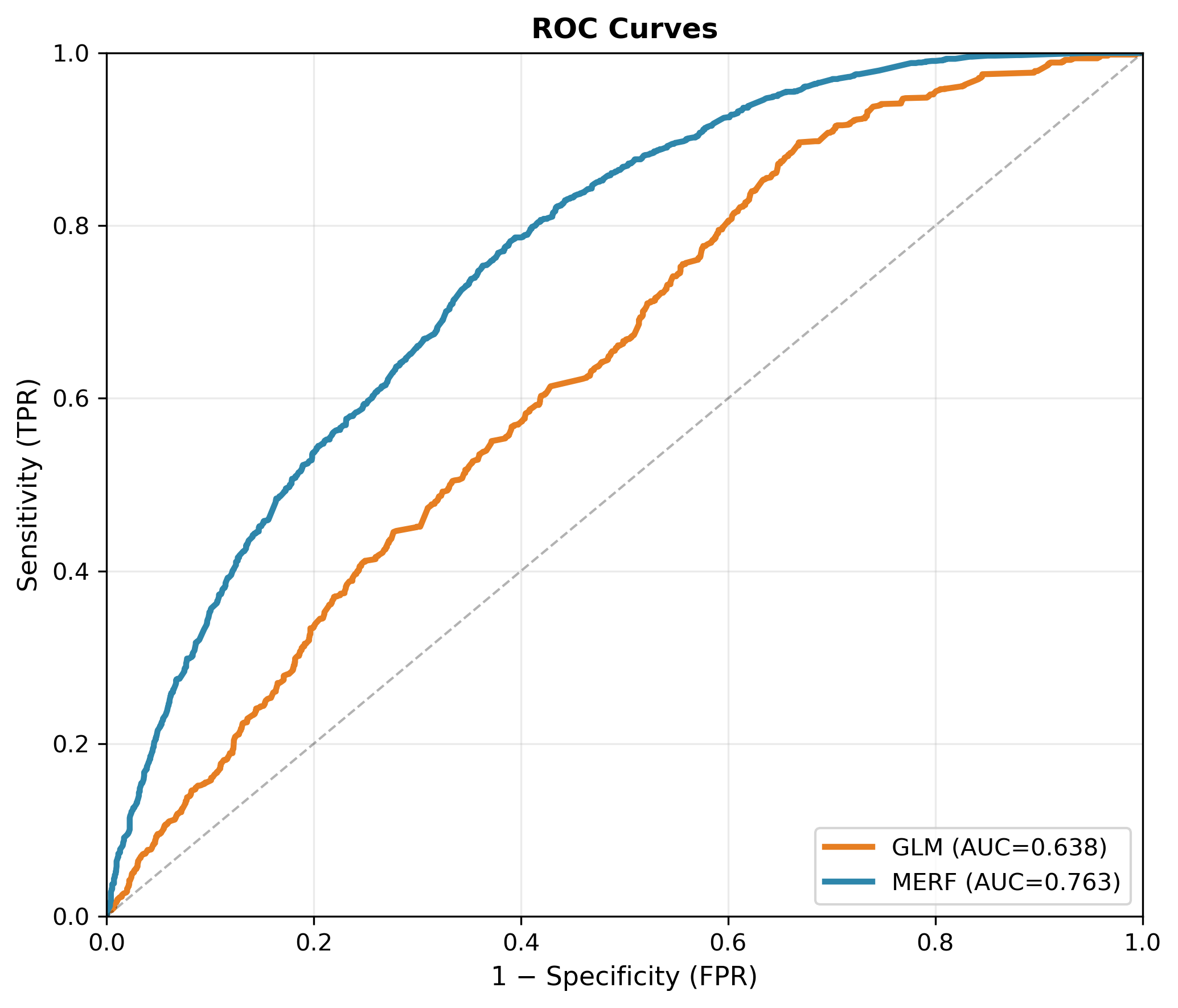}
\caption{ROC curves for the full ten-variable GLM and the full ten-variable MERF on the held-out test set. The MERF curve includes road-specific random effects for roads represented during training.}
\label{fig:roc}
\end{figure}

Figure~\ref{fig:roc} compares the ROC curves for the full ten-variable GLM and the full ten-variable MERF under their originally reported prediction conditions. Results for all model specifications, including the fixed-effects-only MERF evaluation, are reported in Table~\ref{tab:performance}.

\section{Discussion}
\label{sec:discussion}

The null-model ICC indicates that 20.3\,\% of the latent variation in crash severity occurred between roads before the inclusion of explanatory variables, supporting the use of a multilevel framework. This result suggests that crashes recorded on the same road share contextual characteristics that warrant explicit consideration of road-level clustering. The estimated random intercepts further indicate meaningful differences in conditional injury-severity propensity across roads after accounting for the included predictors. The rankings in Table~\ref{tab:roadrankings} may therefore help identify roads for further investigation, but they should not be interpreted as measures of crash frequency or total road risk.

Pavement condition exhibited the largest random-slope variance among the four predictors allowed to vary across roads, followed by lighting condition, driver education, and driver age. Although adverse pavement condition had a negative average coefficient, this association should not be interpreted as evidence that poor pavement improves safety. One possible explanation is that drivers reduce speed under visibly adverse surface conditions, but speed and behavioral responses were not observed in the present data. The random-slope result instead indicates that the association between pavement condition and injury severity differs across roads, potentially because of unmeasured differences in speed, geometry, traffic conditions, maintenance, or driver response in different roads. Similar contextual variation may explain the smaller road-specific differences associated with nighttime crashes.

The predictive comparison separates the contributions of the nonlinear functional form and road-specific information. In the matched ten-variable, fixed-effects-only comparison, MERF achieved an AUC of 0.701, compared with 0.638 for the single-level GLM. This difference indicates that the random-forest component captured predictive structure beyond that represented by the linear logistic specification. Retaining MERF's road-specific random intercepts increased AUC by a further 0.062, to 0.763, reflecting the value of road-specific information when predicting additional crashes on roads represented during training. The full MERF result should therefore not be interpreted as performance for previously unseen roads. The road-level random intercepts were also strongly associated across the two frameworks (Pearson $r=0.892$; Spearman $\rho=0.931$). The slightly higher rank correlation indicates particularly strong agreement in the relative ordering of roads. This consistency reduces concern that the estimated road-level patterns are unique to one functional form, although it does not constitute external validation because both models were fitted using the same data.

Several limitations affect the interpretation of the results. The binary PDO versus non-PDO outcome combines crashes with different injury severities; ordinal or multinomial models could provide a more detailed assessment. The available road-level variables were also limited, and characteristics such as speed limit, lane and shoulder width, curvature, grade, and roadside hazards may explain some of the remaining between-road variation. In addition, the crash-level random split placed crashes from roads represented during training in the test set. The full MERF evaluation, therefore, represents conditional prediction for known roads and may overstate performance for previously unseen roads. Suppressing the MERF random intercepts provides a more comparable assessment of its fixed-effects component, but grouped validation that withholds entire roads would be needed to evaluate generalization to new road contexts. Finally, the random intercepts and SHAP values describe model-based associations rather than causal effects. The road rankings should consequently be treated as a screening tool for prioritizing additional data collection and site assessment, not as definitive evidence that a particular road characteristic causes greater injury severity. The MERF implementation used in this study is a regression-based method originally formulated for continuous outcomes. Its application to the binary 0/1 severity indicator, therefore, provides exploratory predictive scores rather than probabilities from a formally specified binary mixed-effects model. The MERF road-level corrections are consequently not numerically equivalent to the log-odds random intercepts from the multilevel logistic model, and their comparison is limited to association and relative road ordering.

\section{Conclusion}
\label{sec:conclusion}

This study examined road-level heterogeneity in crash severity on rural two-lane highways in Iran. The null-model ICC indicated that 20.3\,\% of the latent variation in severity occurred between roads before explanatory variables were introduced, supporting the use of a multilevel framework. This result suggests that crash outcomes are partly structured by road context and that network-wide average associations may not adequately represent conditions on every road.

Among the predictors allowed to vary across roads, pavement condition had the largest random-slope variance, followed by lighting condition, driver education, and driver age. Although adverse pavement had a negative average association with injury severity, this finding should not be interpreted as evidence that poor pavement reduces risk. Behavioral adaptation, such as reduced speed under adverse surface conditions, is one possible explanation, but it could not be tested with the available data. The road-specific variation indicates that pavement-related associations depend on local conditions that warrant further investigation.

Estimated road-level random intercepts ranged from $-2.87$ to $+1.51$ on the log-odds scale. These estimates identify roads with comparatively higher or lower conditional propensity for a recorded crash to result in a non-PDO outcome after accounting for the included predictors. They may therefore support the prioritization of site investigations, but they do not measure crash frequency, total road risk, or causal effects.

The logistic and MERF random intercepts were strongly associated (Pearson $r=0.892$; Spearman $\rho=0.931$), indicating broadly similar road-level patterns and rankings across the two frameworks. This agreement reduces concern that the ordering arises solely from one model specification, although it does not constitute external validation because both models were fitted using the same data. Overall, the findings support the use of multilevel analysis alongside field assessment to identify and investigate road-specific patterns that may be obscured by single-level, network-wide estimates.

\backmatter

\section*{Declaration of interests}
The authors declare that they have no known competing financial interests or personal relationships that could have appeared to influence the work reported in this paper.

\section*{Funding}
This research did not receive any specific grant from funding agencies in the public, commercial, or not-for-profit sectors.

\section*{Data availability}
The data used in this study are not publicly available due to data-sharing restrictions of the providing agencies.

\section*{Author contributions}
\textbf{Mahdi Azhdari}: Conceptualization, Data curation, Methodology, Software, Formal analysis, Writing -- original draft. \textbf{Ali Tavakoli Kashani}: Supervision, Conceptualization, Writing -- review \& editing. \textbf{Saeideh Amirifar}: Writing -- review \& editing. \textbf{Amirhossein Taheri}:  Methodology, Writing -- review \& editing. \textbf{Gerd M\"{u}ller}: Writing -- review \& editing.

\bibliography{sn-bibliography}

\end{document}